\documentclass[aps,prl,twocolumn,showpacs,superscriptaddress,groupedaddress,nofootinbib,floatfix]{revtex4-1}
   
\usepackage{graphicx,psfrag}
\usepackage{mathrsfs}
\usepackage{amsmath,amsfonts,amssymb}
\usepackage{multirow}
\usepackage{comment}
\usepackage[normalem]{ulem}
\usepackage{hyperref}
\usepackage{xcolor}
\usepackage{acronym}

\newcommand{\reftab}[1]{{Table~(\ref{#1})}}

\newcommand{\reffig}[1]{{Figure~\ref{#1}}}
\newcommand{\refeq}[1]{{Eq.~(\ref{#1})}}

\newcommand{\compressibility}{K_{\rm eq}}
\newcommand{\Kmax}{K_{\rm max}}
\newcommand{\Kmaxtilde}{\tilde{K}_{\rm max}}
\newcommand{\Cmax}{\mathcal{C}_{\rm max}^{\rm TOV}}

\newcommand{\Mmax}{M_{\rm max}^{\rm TOV}}
\newcommand{\Rmax}{R_{\rm max}^{\rm TOV}}
\newcommand{\rhomaxTOV}{n_{\rm max}^{\rm TOV}}
\newcommand{\rhomaxSIM}{n_{\rm max}}
\newcommand{\Msun}{M_{\odot}}
\newcommand{\mthpc}{M_{\rm th}}
\newcommand{\mthpctilde}{\tilde{M}_{\rm th}}

\newacro{NS}{neutron star}
\newacro{EOS}{equation of state}
\newacro{TOV}{Tolman-Oppenheimer-Volkof}
\newacro{GW}{gravitational wave}
\newacro{BNS}{binary neutron star}
\newacro{PC}{prompt collapse}
\newacro{BH}{black hole}
\newacro{MNS}{massive neutron star}
\newacro{EM}{electromagnetic}

\begin{document}

\title{Probing the incompressibility of nuclear matter at ultra-high density through the prompt collapse of asymmetric neutron star binaries}

\author{Albino \surname{Perego}$^{1,2}$\email{email address: albino.perego@unitn.it}}
\author{Domenico \surname{Logoteta}$^{3,4}$}
\author{David \surname{Radice}$^{5,6,7}$}
\author{Sebastiano \surname{Bernuzzi}$^{8}$}
\author{Rahul \surname{Kashyap}$^{5,6}$}
\author{Abhishek \surname{Das}$^{5,6}$}
\author{Surendra \surname{Padamata}$^{5,6}$}
\author{Aviral \surname{Prakash}$^{5,6}$}

\affiliation{${}^{1}$Dipartimento di Fisica, Università di Trento, Via Sommarive 14, 38123 Trento, Italy}
\affiliation{${}^{2}$INFN-TIFPA, Trento Institute for Fundamental Physics and Applications, ViaSommarive 14, I-38123 Trento, Italy}
\affiliation{${}^{3}$Dipartimento di Fisica, Universit\`{a} di Pisa, Largo B.  Pontecorvo, 3 I-56127 Pisa, Italy}
\affiliation{${}^{4}$INFN, Sezione di Pisa, Largo B. Pontecorvo, 3 I-56127 Pisa, Italy}
\affiliation{${}^{5}$Institute for Gravitation \& the Cosmos, The Pennsylvania State University, University Park, PA 16802} 
\affiliation{${}^{6}$Department of Physics, The Pennsylvania State University, University Park, PA 16802}
\affiliation{${}^{7}$Department of Astronomy \& Astrophysics, The Pennsylvania State University,University Park, PA 16802}
\affiliation{${}^8$Theoretisch-Physikalisches Institut, Friedrich-Schiller Universit\"{a}t Jena, 07743, Jena, Germany}

\date{\today}

\begin{abstract}
Using 250 neutron star merger simulations with microphysics, we explore for the first time the role of nuclear incompressibility
in the prompt collapse threshold for binaries with different mass ratios.
We demonstrate that observations of prompt collapse thresholds, either from binaries with two different mass ratios or with one mass ratio but combined with the knowledge of the maximum neutron star mass or compactness, will constrain the incompressibility at the maximum neutron star density, $K_{\rm max}$ to within tens of percent. 
This, otherwise inaccessible, measure of $K_{\rm max}$ can potentially reveal the presence of hyperons or quarks inside neutron stars.
\end{abstract}


\maketitle

\textit{Introduction.-}
The \ac{EOS} of \ac{NS} matter is one of the most fundamental, yet elusive, relations in physics \citep{Lattimer:2015nhk,Oertel:2016bki}. It lays at the interface between several disciplines including nuclear physics, high-energy astrophysics, heavy-ion collisions, multimessenger astronomy and \ac{GW} physics.
Our knowledge of \ac{NS} matter properties is still partial, mostly due to the difficulties in studying strongly interacting bulk matter in the low energy limit typical of nuclear interactions \citep{Machleidt:2011zz}.
Even the appropriate degrees of freedom are uncertain: while
nucleons are the relevant species around the nuclear saturation density,
$n_0 = 0.16~{\rm fm^{-3}}$, it is still unclear if hyperons \cite{Chatterjee:2015pua,Logoteta:2021iuy} or a phase transition to quark matter \cite{Bombaci:2016xuj,2009RvMP...81.1031B,Benic:2014jia} can appear at densities $n\gtrsim 2n_0$ in \ac{NS} interiors.

\ac{NS} \ac{EOS} models are experimentally constrained by the masses of ordinary
nuclei, as well as by the energy per baryon and its derivatives with respect to baryon density, $n_b$, around $n_0$ and close to isospin symmetry, i.e. for symmetry parameter $\delta \equiv (n_n - n_p)/n_b \approx 0$, $n_{n,p}$ being the density of neutrons and protons. If $P$ is the matter pressure, the nuclear incompressibility of cold nuclear matter at fixed composition is defined as
\begin{equation}
    K(n_b,\delta) \equiv \left. 9 \frac{\partial P}{\partial n_b} \right|_{T=0,\delta={\rm const}} \, .
\label{eq:incompressibility}
\end{equation}
It describes the response of matter to compression and its value 
can be currently measured only for symmetric matter at saturation density, $K_{\rm sat}$, although with some controversy \cite{Garg:2018uam, 2006EPJA...30...23S, Youngblood:1999zza, Stone:2014wza, Avogadro:2013tza}. 
While isoscalar giant monopole resonance experiments for closed shell nuclei
provided $K_{\rm sat} = (240 \pm 20) {\rm MeV}$,
studies based on open-shell nuclei reported quite different values in the range $250 {\rm MeV}$-$315 {\rm MeV}$ \cite{Stone:2014wza} or even values around $200 {\rm MeV}$ \cite{Avogadro:2013tza}.
Nevertheless, $K_{\rm sat}$ is unconstrained at densities and compositions relevant for \acp{NS}
(far from $n \approx n_0$ and $\delta \approx 0$). 
In particular, the \ac{NS} central density increases monotonically with the \ac{NS} mass and at the stability limit, corresponding to mass and radius ($\Mmax$, $\Rmax$), can reach $\rhomaxTOV \sim 4-7 n_0$, depending on the \ac{EOS}.
Moreover, for $n_b \gtrsim n_0$, $\beta$-equilibrated matter is very neutron rich, $\delta_{\rm eq} \sim 1$.

In addition to nuclear constraints, astrophysical \ac{NS} properties
provide useful insights on the \ac{EOS}.
Constraints derived from the observation of massive, isolated \ac{NS} \citep{Demorest:2010bx,Antoniadis:2013pzd,NANOGrav:2019jur,Fonseca:2021wxt,Miller:2021qha,Riley:2021pdl,Zhang:2021xdt}, from \ac{GW} signals \citep{Abbott:2018exr,De:2018uhw} and multimessenger observations of \ac{BNS} mergers \citep{Radice:2017lry,Radice:2018ozg,Bauswein:2017vtn,Margalit:2017dij,Most:2018hfd,Breschi:2021tbm},
or by their combination \citep{Raaijmakers:2021uju,Pang:2021jta}, are very informative about the high density regime.
A key phenomenon in this respect is the \ac{PC} to \ac{BH} of the merger remnant, since this behavior can influence both the \ac{GW} and \ac{EM} signals produced by \ac{BNS} mergers \citep{Hotokezaka:2012ze,Hotokezaka:2013iia,Bauswein:2013yna,Agathos:2019sah,Radice:2018pdn,Bernuzzi:2020txg}.
The \ac{PC} behavior of equal mass \acp{BNS} was extensively explored in Ref.~\citep{Shibata:2005ss,Hotokezaka:2011dh,Bauswein:2013jpa,Koppel:2019pys,Bauswein:2019ybt,Bauswein:2020aag,Kashyap:2021wzs}.
It was shown, for example, that the threshold mass for \ac{PC}, $\mthpc$, normalized to $\Mmax$, linearly correlates with the maximum compactness, defined as $\Cmax \equiv G\Mmax/(\Rmax c^2)$, where $c$ and $G$ are the speed of light and the gravitational constant, respectively,
as well as with other \ac{NS} equilibrium properties.
More recently, also the study of asymmetric \ac{BNS} mergers has received attention \citep{Bauswein:2020aag,Bauswein:2020xlt,Tootle:2021umi,Kolsch:2021lub}.
\citet{Bauswein:2020aag,Bauswein:2020xlt} concluded that \ac{PC} in asymmetric BNS usually occurs for masses equal or smaller than in the equal mass case, with the possible exception of modest asymmetries and very soft \acp{EOS}. The total mass reduction is stronger for more extreme mass ratios and it has a non-trivial dependence on the \ac{NS} \ac{EOS}.
\citet{Tootle:2021umi} suggested instead a quasi-universal relation. In all these works, several fitting formulae to numerical results were provided.

In this \textit{Letter}, we show that $K_{\max}$, the incompressibility of nuclear, $\beta$-equilibrated matter at $\rhomaxTOV$, 
determines the behavior of \ac{BNS} mergers close to \ac{PC} 
and, in particular, their dependence on the mass ratio.
Our results stem from the largest set of numerical relativity simulations of irrotational, asymmetric binaries with finite temperature, composition dependent microphysical \acp{EOS} to date. 
We demonstrate that the detection of $\mthpc$ at two different mass ratios can provide a direct measurement of $\Kmax$ in a regime otherwise inaccessible. Additionally, we suggest that its value can yield information about the relevant thermodynamics degrees of freedom close to $\rhomaxTOV$.

\textit{Methods and models.}- 
We simulate 250 irrotational \ac{BNS} mergers with different gravitational masses $M \equiv M_1 + M_2 \in [2.786 \Msun,3.3\Msun]$ and mass ratios $q \equiv M_1/M_2 \in \left\{ 0.6,0.65,0.7,0.75,0.85,1 \right\}$.
We perform series of simulations at fixed $q$ 
while changing $M$ to explore the onset of the \ac{PC} behavior
and determine $\mthpc(q)$.
For the definition of $\mthpc(q)$ and its numerical error, $\delta \mthpc(q)$, we follow Refs.~\citep{Bauswein:2012ya,Bauswein:2020aag,Kashyap:2021wzs},
monitoring the maximum of the rest mass density, $\rhomaxSIM$, throughout the computational domain. 
Simulations are performed with the same codes and
setup as in Ref.~\cite{Kashyap:2021wzs}; $q=1$ data are from  Ref.~\cite{Kashyap:2021wzs}, while $q\neq1$ data are presented here for the first time. See the Supplemental Material for more details.

\begin{figure}[t]
\centering
\includegraphics[width=\linewidth]{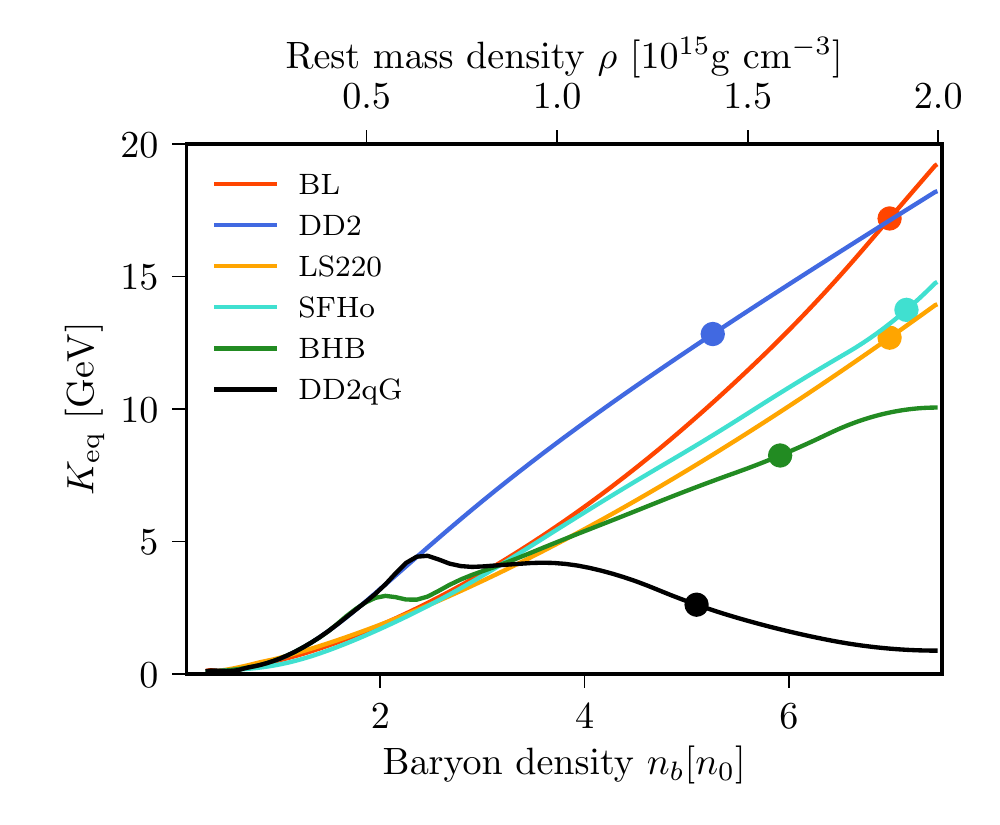}
\caption{Nuclear incompressibility $\compressibility$ of cold, $\beta$-equilibrated nuclear matter as a function of baryon density for the \acp{EOS} employed in this work. Solid markers correspond to $\Kmax$, i.e. $\compressibility$ at the central density of the heaviest, irrotational \ac{NS}.}
\label{fig: compressibility}
\end{figure}

To span present uncertainties, we consider six finite-temperature, composition dependent \ac{NS} \acp{EOS}.
Four are purely nucleonic and widely used:
BL \citep{Bombaci:2018ksa,Logoteta:2020yxf}, 
SFHo \citep{Steiner:2012rk} and HS(DD2) \citep[][hereafter DD2]{Typel:2009sy,Hempel:2009mc}, and LS220 \citep{Lattimer:1991nc}. 
Additionally, we consider an EOS including hyperons, HS(BHB$\Lambda\phi$) \citep[][hereafter BHB]{Banik:2014qja}, and one including a phase transition to quark matter, DD2qG, also presented in Ref.~\citep{Kashyap:2021wzs}. In both cases, the nucleonic baseline is DD2.
In \reffig{fig: compressibility}, we present the nuclear incompressibility of neutrinoless, $\beta$-equilibrated, cold \ac{NS} matter, $K_{\rm eq}$, defined as in \refeq{eq:incompressibility} but for $\delta=\delta_{\rm eq}$,
for the six different EOSs above as a function of $n_b$. For each \ac{EOS} we highlight $\Kmax \equiv K(\rhomaxTOV)$. It is striking that the properties in the low density regime ($n_b < 2n_0$) do not necessarily correlate with those at $n_b \sim \rhomaxTOV$. Moreover, the BL, SFHo and LS220 \acp{EOS}, despite being softer than the DD2 \ac{EOS}, reach larger $\rhomaxTOV$ and provide similar, if not larger, $\Kmax$. 

\begin{figure}[t]
\centering
\includegraphics[width=\linewidth]{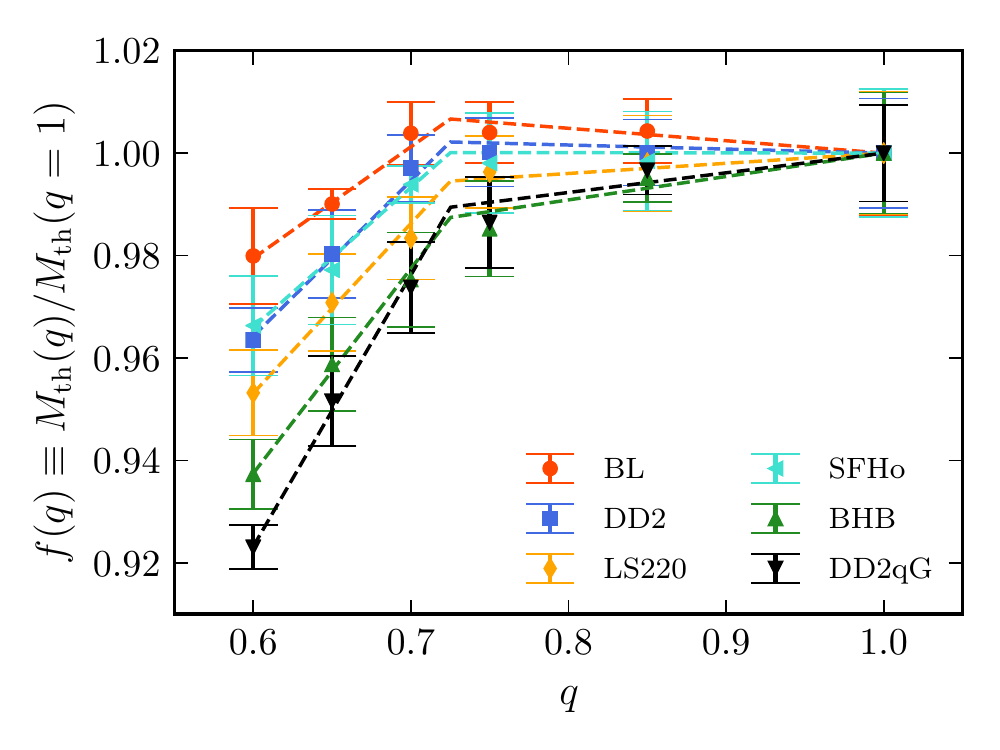}
\caption{Threshold \ac{PC} masses normalized to the $q=1$ case as a function of $q$ for all the \ac{EOS} used in this work. Dashed lines correspond to \refeq{eq: double linear fit} fit.} 
\label{fig: bilinear fit without residuals}
\end{figure}

\textit{Results}.- Our simulations robustly indicate that \ac{PC} occurs as $\rhomaxSIM$ approaches $\rhomaxTOV$ at merger. With the exception of the DD2qG \ac{EOS}, for which $\rhomaxSIM \sim 0.8-1.2 \rhomaxTOV$,
for the heaviest non-\ac{PC} \ac{BNS} we observe $\rhomaxSIM\sim 0.75-0.95 \rhomaxTOV$ at the first remnant bounce, with larger values usually associated to $q \sim 1$.

Two opposite effects influence the evolution of $\rhomaxSIM$ with respect to $q$.
On the one hand, for a given $M$, binaries with smaller $q$'s have smaller orbital angular momentum and the \ac{NS} cores are more prone to fuse (and thus to increase $\rhomaxSIM$ toward $\rhomaxTOV$) due to the smaller rotational support \citep{Bauswein:2020aag,Bauswein:2020xlt}. 
On the other hand, the nuclear incompressibility usually increases as $\rhomaxSIM$ grows, providing a larger nuclear repulsion that contrasts its further increase.
Since \ac{PC} is observed for $\rhomaxSIM \sim \rhomaxTOV$, it is understandable that $\Kmax$ is the incompressibility value relevant for the PC behavior.

To analyse the dependence of \ac{PC} on $\Kmax$,
in \reffig{fig: bilinear fit without residuals} we first consider 
$f(q) \equiv \mthpc (q)/ \mthpc (q=1)$ for all EOSs, where (to be conservative) the error bars have been obtained by
propagating the errors both on $M_{\rm th}(q)$ and  $M_{\rm th}(q=1)$. 
Values of $\mthpc$ and $\delta \mthpc$ are reported in the Supplemental Material.
We first observe that our results do not have a universal behavior for the different \acp{EOS}. 
Second, we notice that a variation of almost a factor of 1.7 in $q$ has a small effect on $\mthpc(q)$, with the corresponding variation in $f(q)$ ranging between 3\% and 8\%, larger for \acp{EOS} with a smaller $\Kmax$. 
This is broadly compatible to what observed in \citep{Bauswein:2020aag,Bauswein:2020xlt,Tootle:2021umi,Kolsch:2021lub}
and should be compared with the larger ($\lesssim 20\%$) variation in $\Mmax$ or $\mthpc (q=1)/\Mmax $ reported in Refs.~ \citep{Hotokezaka:2011dh,Bauswein:2012ya,Koppel:2019pys,Agathos:2019sah,Bauswein:2019ybt,Bauswein:2020aag,Kashyap:2021wzs}.
 
Focusing on the behavior of $f(q)$ for $ 0.7 \lesssim q \leq 1$
we observe that, depending on the \ac{EOS}, $f(q)$ can decrease,
stay approximately constant or even increase as $q$ decreases \citep[see also][]{Kiuchi:2019kzt,Bauswein:2020xlt,Kolsch:2021lub}.
We interpret this as the result of the interplay between the binary orbital angular momentum 
and the incompressibility of nuclear matter, in light of the merger dynamics.
For \acp{BNS} with $q \lesssim 1$ and $M \approx \mthpc$, the central density inside the more massive \ac{NS} ranges in 0.40-0.49~$\rhomaxSIM$ (depending on the \ac{EOS}) and the merger is driven by the fusion of two comparable \ac{NS} cores.
If $\compressibility$ increases steeply enough with $\rhomaxSIM$, nuclear repulsion contrasts efficiently gravity-driven compression.
The net result is that for \ac{EOS}s with a relatively large $\Kmax$ (as BL, SFHo and DD2), $M_{\rm th}( q \lesssim 1 )$ can stay rather constant or even increase as $q$ decreases. On the opposite, if $\compressibility$ does not increase significantly with $n_b$ and $\Kmax$ is relatively low (as for DD2qG and BHB),
nuclear repulsion is not enough to counterbalance the lack of rotational support and \ac{PC} occurs for $M_{\rm th}( q \lesssim 1 ) < M_{\rm th}( q=1 )$.

Moving to $ 0.6 \lesssim q \lesssim 0.7$, we notice a clear change of behavior:
$f(q)$  decreases as $q$ decreases for all \acp{EOS}.
But, once again, the variation  depends sensitively on $\Kmax$:
\acp{EOS} characterized by a smaller $\Kmax$ result not only in smaller $f(q)$, but also in larger relative variations with respect to $f(q \approx 0.7)$.
We explain this transition in terms of the different merger dynamics. For \acp{BNS} at the \ac{PC} threshold and with $q \lesssim 0.7$, the central density inside the more massive \ac{NS} increases to 0.5-0.57~$\rhomaxTOV$, while the secondary \ac{NS} is more significantly deformed and tidally disrupted during the last orbits.
As $q$ decreases, the denser core of the more massive \ac{NS} is compressed by more massive streams of accreting matter \citep{Bauswein:2013yna,Dietrich:2015pxa,Dietrich:2016hky,Bernuzzi:2020txg,Bauswein:2020aag}.
The nuclear incompressibility still opposes compression, but less efficiently than in the $0.7 \lesssim q \leq 1$ regime.
$\Kmax$ still provides a measure of the \ac{NS} matter resistance to compression in the relevant density interval and different \acp{EOS} result in different relative variations.

Our data qualitatively agree with those from independent simulations recently reported in Refs.~\citep{Bauswein:2019ybt,Tootle:2021umi,Kolsch:2021lub}. However, quantitative differences comparable to the overall variation observed in our results are found. This is possibly due to different definitions of \ac{PC} threshold, gravity treatment or numerical resolutions. A comparison with the some of the available fits is reported in the Supplemental Material. Moreover, our extended set of EOS indicates a sub-leading but significant and systematic \ac{EOS} dependence emerging for asymmetric binaries, in contrast to a quasi-universal behavior~\citep{Tootle:2021umi}.

\reffig{fig: bilinear fit without residuals} suggests the existence of two different regimes, separated 
by $0.7 \lesssim \tilde{q} \lesssim 0.75$, which is largely independent from the \ac{EOS}. 
In each of the two regimes, $f(q)$ is well described by a linear relation. Thus, for each \ac{EOS} we fit our data by considering:
\begin{equation}
f(q) = \alpha(q) q + \beta(q) =
    \begin{cases}
    \alpha_l q + \beta_l & {\rm if}~q < \tilde{q} \, , \\
    \alpha_h q + \beta_h & {\rm if}~q \geq \tilde{q}    \, .
    \end{cases}
    \label{eq: double linear fit}
\end{equation}
We fix $\beta_{l,h}$ in \refeq{eq: double linear fit} by imposing 
the continuity of $f(q)$ at $q=\tilde{q}$ and 
$f(q=1)=1$.
Moreover, we assume $\tilde{q} = 0.725$ by closely inspecting \reffig{fig: bilinear fit without residuals}.
Least square fits (dashed lines) are performed on the two parameters $\alpha_{l,h}$, corresponding to the slopes of the two linear regimes.
The residuals relative 
to the errors are always smaller than 0.5 and without clear
systematic trends both with respect to the \ac{EOS} and $q$.

\begin{figure}[t!]
\centering
\includegraphics[width=\linewidth]{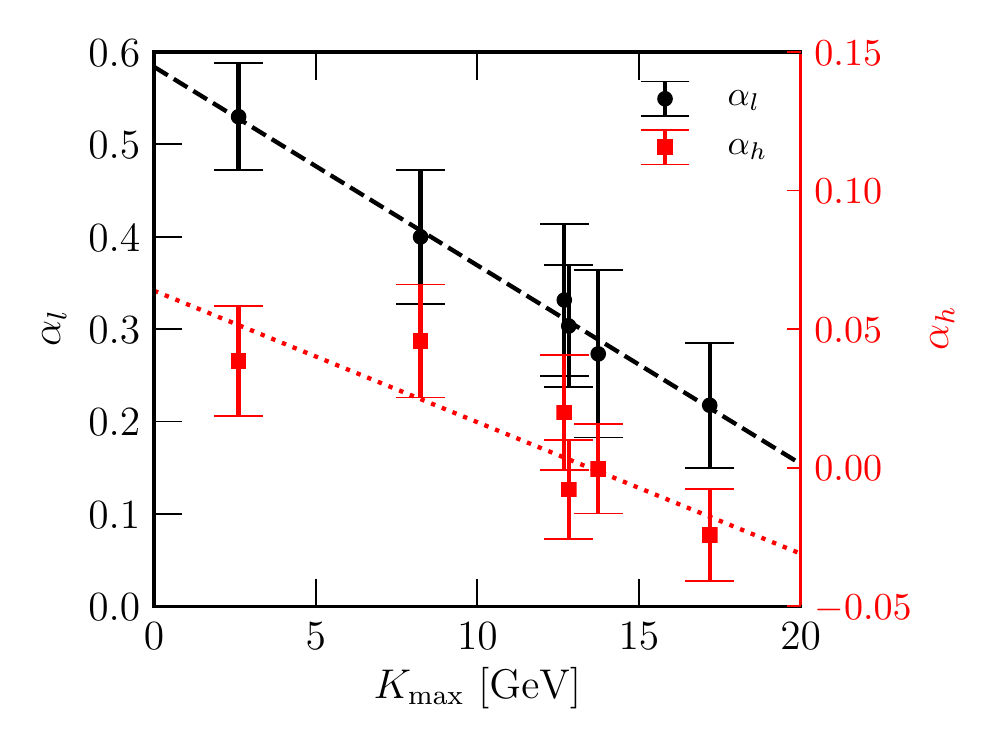}
\caption{Slopes of the fitting coefficients $\alpha_{l,h}$ for data in \reffig{fig: bilinear fit without residuals} as a function of $\Kmax$. Both slopes show a trend with $\Kmax$ that we fitted with a linear function (dashed lines).}
\label{fig: slope correlation}
\end{figure}

Our simulations reveal a correlation between $\alpha_{l,h}$ and $\Kmax$ supporting the interpretation that the latter is one of the key properties that control the \ac{PC}.
In \reffig{fig: slope correlation}, we represent $\alpha_{l,h}$ 
with their uncertainties as a function of $\Kmax$ for each \ac{EOS}. 
Given the reduced number of \acp{EOS} and the relatively large
uncertainties, we fit $\alpha_{l,h}$ with a first order polynomial in $\Kmax$ (dashed lines in \reffig{fig: slope correlation}):
\begin{eqnarray}
    \alpha_l &=& - (22 \pm 1) {\rm TeV^{-1}} \Kmax + (0.58 \pm 0.01)  \, , \nonumber \\
    \alpha_h &=& - (4.7 \pm 1.0) {\rm TeV^{-1}} \Kmax + (0.064 \pm 0.017)  \, .
    \label{eq: fit coefficients}
\end{eqnarray}
The slopes of the linear behaviors observed in \reffig{fig: bilinear fit without residuals}
usually decrease as the incompressibility increases. This
confirms that \acp{EOS} with a large incompressibility provide a possible increase in $M_{\rm th}$ for $ \tilde{q} < q \leq 1$ and a less steep decrease for $q < \tilde{q} $.

\begin{figure}[t]
\centering
\includegraphics[width=1.\linewidth]{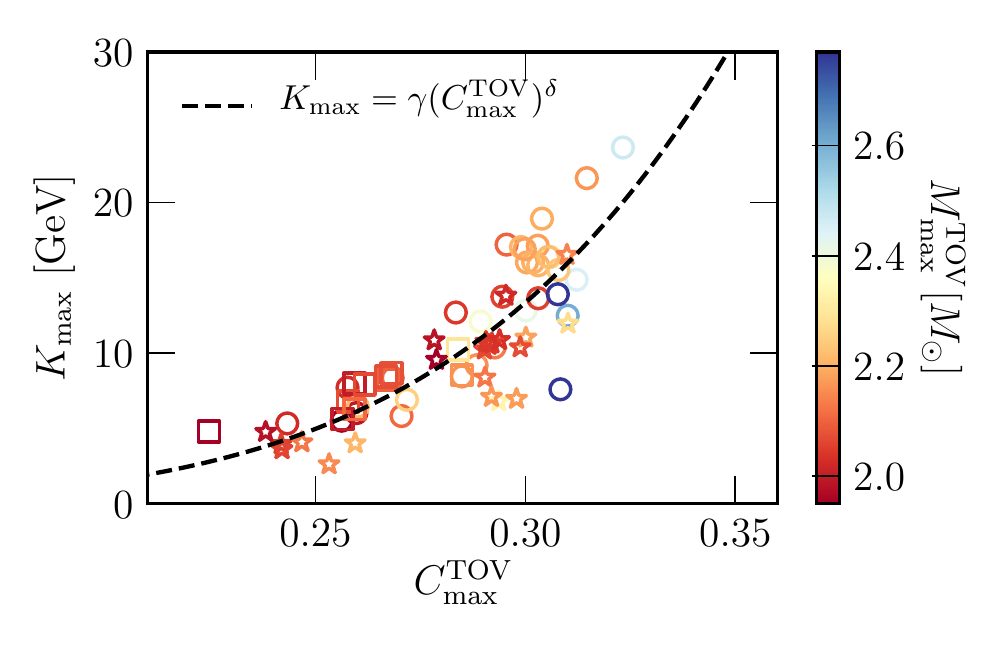}
\caption{$\Kmax$
as a function of the compactness of the heaviest \ac{NS} for a large sample of \acp{EOS}. 
Circles correspond to nucleonic \acp{EOS}, while squares and stars to \acp{EOS} containing hyperons or undergoing a phase transition to quarks, respectively. 
}
\label{fig: all incompressibilities}
\end{figure}


\textit{Discussion.-} Our results suggest that the determination of $\mthpc$ at two different $q$'s, $q_{1,2}$, allows to determine $\Kmax$ by solving:
\begin{equation}
    \frac{\mthpc(q_1)}{\mthpc(q_2)} =
    \frac{\alpha(\Kmax,q_1)q_1+\beta(\alpha)}{\alpha(\Kmax,q_2)q_2+\beta(\alpha)} \, ,
    \label{eq: K eq for two q}
\end{equation}
where $\alpha$ and $\beta$ are defined consistently with Eqs.~(\ref{eq: double linear fit}) and (\ref{eq: fit coefficients}).
To test this, we repeat the previous fits excluding results from the SFHo \ac{EOS}. The new fitted coefficients, $\alpha'_{l,h}$, are compatible to within uncertainties with $\alpha_{l,h}$ in \refeq{eq: fit coefficients}.
We deduce $\Kmax$ for SFHo using these new fits and the $M_{\rm th}(q)$ SFHo results at two different $q$'s.
In particular, we randomly sample the intervals $(\mthpc(q) \pm \delta \mthpc(q)/2)$ one thousand times to set simulated values for the threshold masses, $\mthpctilde(q_{1,2})$, and to compute $\Kmaxtilde$
by solving  \refeq{eq: K eq for two q}.  
We finally extract the average relative discrepancy between the computed and actual values.
For example, using $M_{\rm th}(q=0.65)$ and $M_{\rm th}(q=0.85)$ we recover $\Kmax$ to within 2\% of its actual value.
The uncertainty increases when considering $M_{\rm th}(q=0.7)$ and $M_{\rm th}(q=0.85)$.
In this case, $\Kmax$ is recovered to within 15\%.
Our method does not necessarily require the knowledge of $\mthpc(q)$ at two $q \neq 1$.
For example, using $\mthpc(q=0.7)$ and $\mthpc(q=1)$ 
we recover $\Kmax$ within 3.5\%.
The above discrepancies on $\Kmax$ are compatible with the uncertainties implied by \reffig{fig: slope correlation}.

To further challenge our method, we consider the independent results for $\mthpc(q)$ from Ref.~\cite{Tootle:2021umi} obtained for irrotational \acp{NS} and for the TNTYST EOS \cite{Togashi:2017mjp}, an \ac{EOS} not included in our sample 
and for which $\Kmax > 20 {\rm GeV}$ \footnote{We notice, however, that the TNTYST \ac{EOS} becomes acausal close to $\rhomaxTOV$.}. We consider the $\alpha_{l,h}$ fits, \refeq{eq: fit coefficients}, and we solve Eqs.~(\ref{eq: K eq for two q}) and (\ref{eq: inverse power law fit}) using $M_{\rm th}(q=0.7)$ and $M_{\rm th}(q=0.9)$.
Despite possible systematical differences related to the different way to 
determine $\mthpc$, we recover the expected value of $\Kmax$ 
to within 25\% ($\sim$6 GeV) of its actual value.

Prompted by these results, we investigate a direct correlation between $\Cmax$ and $\Kmax$ and we find that the values of $\Kmax$
can provide information on the relevant degrees of freedom in ultradense matter.
In \reffig{fig: all incompressibilities}, we present $\Kmax$ as a function of $\Cmax$ for a large set of EOSs.
In particular, we selected EOSs that stay causal up to $\rhomaxTOV$ and for which $\Mmax > 1.97 \Msun$. More detailed information can be found in the Supplemental Material.
Different symbols refer to different particle contents 
while colors to $\Mmax$. We suggest that
large  $\Kmax (\gtrsim 15$GeV) are more easily associated with purely nucleonic \acp{EOS}, while \acp{EOS} containing hyperons or showing a phase transition to quarks are characterized by
small $\Kmax (\lesssim 15{\rm GeV})$.
A tighter threshold at 12GeV can be observed 
if only two \acp{EOS} containing just $u$ and $d$ quarks, out a sample of 34 \acp{EOS} containing quarks or hyperons, were removed.
Moreover, $\Kmax$ can be fitted in good approximation with a power law, $\Kmax = \gamma(\Cmax)^\delta$. Standard least squared methods provide
$\gamma = (9.2 \pm 5.4){\rm TeV~}$ and $ \delta = 5.67 \pm 0.50$.
Despite not being trivial, such a relation is not surprising, since both
$\Mmax$ and $\Rmax$ depend on the equilibrium response of the heaviest \ac{NS} 
to radial perturbations for $n_b \sim \rhomaxTOV$, and thus on $\Kmax$. Moreover, it provides a possible connection between our findings
and previous, different fits for $\mthpc$ expressed in terms 
of $\Mmax$ and $\Cmax$, both for symmetric and asymmetric mergers \citep{Bauswein:2013jpa,Koppel:2019pys,Bauswein:2020aag, Bauswein:2020xlt,Kashyap:2021wzs,Kolsch:2021lub}. 
For example, we have repeated our analysis in terms of $\Cmax$ rather 
than $\Kmax$, finding comparable results, as reported in the Supplemental Material. Even if this relation directly connects $\Kmax$ to $\Cmax$, we stress that $\Kmax$ 
provides a cleaner and more intuitive physical interpretation of the \ac{PC} behavior for $q \neq 1$.

This $\Kmax(\Cmax)$ relation, combined with 
the linear relation $\mthpc(q=1)/\Mmax = a \Cmax + b$ first proposed in Ref.~\citep{Bauswein:2013jpa} but with coefficients from Ref.~\cite{Kashyap:2021wzs}, suggests that 
$\Mmax$ can be also related to $\Kmax$ and $\mthpc(q=1)$:
\begin{equation}
 \Mmax = \frac{\mthpc(q=1)}{a (\Kmax/ \gamma)^{1/\delta} + b} \, .
 \label{eq: inverse power law fit}
\end{equation}
Eqs.~(\ref{eq: double linear fit}) and (\ref{eq: inverse power law fit}) together suggest that $\Kmax$ 
can be estimated by the knowledge of only one $M_{\rm th}(q)$, if $\Mmax$ is known:
\begin{equation}
    \frac{\mthpc(q)}{\Mmax(a (\Kmax/\gamma)^{1/\delta}+b)} = \alpha(\Kmax,q) q + \beta(\alpha) \, .
\label{eq:implicit eq for K}    
\end{equation}
For example, using the $\alpha'_{l,h}$ fits while employing $\Mmax$ and $\mthpc(q=0.85)$ SFHo results as input data, we recover $\Kmax$ and $\Cmax$ to within 40\% and 1.6\%, respectively.
Comparable results were obtained from smaller $q$'s.

The analogy between the definition of $K_{\rm eq}$ and the square of the speed of sound of \ac{NS} matter, $c_{\rm s}^2 = \left. \partial P/\partial \epsilon \right|_{T=0,\delta_{\rm eq}}$, where $\epsilon$ is the density of internal energy, suggests that the measurement of the \ac{PC} threshold at different $q$'s can also provide constraints on the value of $c_{\rm s}^2$ close to $\rhomaxTOV$. Indeed, the $\alpha$ and $\beta$ coefficients of \refeq{eq: double linear fit} also correlate with $c_{\rm s}^2$ in a comparable way as with $\Kmax$ and $\Cmax$, as visible in the Supplemental Material. Constraints on $c_{\rm s}$ can provide further insight into the physics governing the \ac{EOS} of nuclear matter \citep[see e.g.][]{Tews:2018kmu,Capano:2019eae}.

The larger detection horizon associated to massive \ac{BNS} mergers suggests that, as in the case of GW190425 \citep{Abbott:2020uma}, \acp{PC} are a viable observational phenomenon associated to a significant fraction of \acp{BNS} that will become accessible in the next \ac{GW} observing runs \citep{Aasi:2013wya,KAGRA:2013rdx} and with 3rd generation \ac{GW} detectors \citep{Punturo:2010zz,Maggiore:2019uih,2019BAAS...51g..35R}.
While current \ac{GW} detections allow the precise measurement of the chirp mass and, up to a certain extent, of the total mass, the mass ratio is more uncertain. High enough signal-to-noise ratios and good sky localizations favoring followup \ac{EM} observations will be key to provide better constraints on $q$.
We estimate the possible impact of the uncertainties on $\mthpc$ and on $q$ on the estimate of $\Kmax$ by solving again \refeq{eq: K eq for two q}, using the $\alpha'_{l,h}$ fitted coefficients (i.e. considering SFHo as our underlying EOS and removing it from our fitting sample). 
We randomly sample both $\mthpc$ and $q$ within $\mthpc \pm \Delta \mthpc$ and $q \pm \Delta q$, where $\Delta \mthpc$ and 
$\Delta q$ are the uncertainties in the determination of $\mthpc(q)$. 
In the case of $q=0.85$ and $q=0.65$, to determine $\Kmax$ with at least 30\% accuracy at 90\% confidence level we estimate $\Delta \mthpc \lesssim 0.025 \Msun$ and 
$\Delta q \lesssim 0.05$.
For $q=0.85$ and $q=0.70$, the uncertainties should decrease to $\Delta \mthpc \lesssim 0.01 \Msun$ and $\Delta q \lesssim 0.025$ to get a similar accuracy. The difference between the two cases proves that, due to the larger slope of $\mthpc(q)$ at $q \ll 1$, the determination of $\mthpc(q)$ for very asymmetric systems is more constraining. Such uncertainties are within reach of future observations and detectors \cite{Borhanian:2022czq}.

More theoretical \ac{PC} studies will be needed to reduce systematic uncertainties and include more detailed physics. Nevertheless, our results clearly indicate a new and unique way to access critical information on extreme density nuclear physics using observations of promptly collapsing \ac{BNS} mergers.

\begin{acknowledgments}
\textit{Acknowledgments.-} 
A.Pe. thanks Matteo Breschi for useful discussions. A.Pe., D.L. and S.B. acknowledge the INFN for the usage of computing and storage resources through the \texttt{tullio} cluster in Turin. 
The authors acknowledge the usage of \ac{EOS} tables from the CompOSE website, https://compose.obspm.fr. D.L. thanks also C. Providencia for providing some \ac{EOS} tables.  
A.Pe. and D.L. acknowledge PRACE for awarding them access to Joliot-Curie at GENCI@CEA. A.Pe. also 
acknowledges the usage of computer resources under a CINECA-INFN agreement (allocation INF20\_teongrav and INF21\_teongrav).  
S.B. acknowledges funding from the EU H2020 under ERC Starting Grant,
no.BinGraSp-714626, and from the Deutsche Forschungsgemeinschaft, DFG,
project MEMI number BE 6301/2-1. 
D.R. acknowledges funding from the U.S. Department of Energy, Office of Science, Division of Nuclear Physics under Award Number(s) DE-SC0021177 and from the National Science Foundation under
Grants No. PHY-2011725, PHY-2020275, PHY-2116686, and AST-2108467.
NR simulations were performed on Joliot-Curie at GENCI@CEA (PRACE-ra5202),
SuperMUC-LRZ (Gauss project pn56zo), Marconi-CINECA (ISCRA-B project HP10BMHFQQ, INF20\_teongrav and INF21\_teongrav allocation); Bridges, Comet, Stampede2 (NSF XSEDE allocation TG-PHY160025), NSF/NCSA Blue Waters (NSF AWD-1811236), 
supercomputers. This research used resources of the National Energy Research
Scientific Computing Center, a DOE Office of Science User Facility supported by
the Office of Science of the U.S.~Department of Energy under Contract
No.~DE-AC02-05CH11231.
\end{acknowledgments}

\section{Supplemental Material}

\subsection{Simulation sample}

We evolve our \acp{BNS} using the publicly available \texttt{WhiskyTHC} code 
\cite{Radice:2012cu, Radice:2013hxh, Radice:2013xpa}, built on the top of the 
\texttt{EinsteinToolkit} \cite{Loffler:2011ay}. \texttt{WhiskyTHC} solves general 
relativistic hydrodynamics in conservative form using high-resolution shock-capturing finite-volume schemes based on high-order reconstruction operators.
The spacetime metric is evolved in the Z4c formulation of Einstein's equations
\cite{Bernuzzi:2009ex, Hilditch:2012fp} implemented by the \texttt{CTGamma} module 
\cite{Pollney:2009yz, Reisswig:2013sqa} of \texttt{EinsteinToolkit}. 

To properly resolve the merger dynamics, we employ an adaptive mesh refinement consisting in seven nested grids with 1:2 linear scaling between consecutive 
grids and provided by the \texttt{Carpet} library \cite{Schnetter:2003rb}. The latter implements the Berger-Oliger scheme with refluxing \cite{Berger:1984zza,Berger:1989a}.
The resolution of the innermost grid, $h$, characterizes each simulation and we distinguish between low ($h \approx 246 \; \rm{m}$, LR), standard ($h \approx 185 \; \rm m$, SR) and high ($h \approx 123 \; \rm{m}$, HR) resolutions.
More details on the grid setup were discussed in detail in Ref.~\cite{Radice:2018pdn}. All simulations were performed with microphysical EOS, also accounting for neutrino emission using the leakage scheme discussed in \cite{Galeazzi:2013mia, Radice:2016dwd}.

Initial conditions are obtained by the pseudospectral elliptic solver \texttt{Lorene} \cite{Gourgoulhon:2000nn}. The initial separation between the centers of the two stars is typically taken to be $40\ {\rm km}$.

All configuration immediately above or below $M_{\rm th}$ were run at least at two 
resolutions (LR and SR). For a few selected cases we performed also HR runs. 
In total, we performed 250 BNS merger simulations: 125 at LR, 116 at SR and 9 at HR. 
The \ac{PC} threshold and its error have been computed using the two highest 
available resolutions. 
Thus the error bars include numerical resolution uncertainty. 
A summary of our results is reported in \reftab{table: PC results}.

\begin{table*}[h!]
\begin{center} 
\caption{Summary of the threshold masses and of their error of numerical origin, for each of the equations of state used in this work and for all the explored mass ratios.}
\label{table: PC results} 
\begin{tabular}{ | c | c | c | c | c | c | c |  } 
 \hline\hline 
 EOS   &  \multicolumn{6}{c|}{$(\mthpc \pm \delta \mthpc)~[\Msun]$}  \\ \hline
    & $q=0.6$ & $q=0.65$ & $q=0.7$ & $q=0.75$ & $q=0.85$ & $q=1.0$ \\
BL & $2.865 \pm 0.019$ & $2.895 \pm 0.006$ & $2.935 \pm 0.012$ & $2.936 \pm 0.012$ & $2.937 \pm 0.012$ & $2.924 \pm 0.024$  \\ 
DD2   & $3.155 \pm 0.014$ & $3.210 \pm 0.020$ & $3.264 \pm 0.015$ & $3.275 \pm 0.015$ & $3.274 \pm 0.014$ & $3.274 \pm 0.024$ \\
LS220 & $2.817 \pm 0.017$ & $2.870 \pm 0.020$ & $2.907 \pm 0.017$ & $2.945 \pm 0.014$ & $2.950 \pm 0.019$ & $2.956 \pm 0.025$ \\
SFHo  & $2.729 \pm 0.019$ & $2.760 \pm 0.021$ & $2.807 \pm 0.007$ & $2.819 \pm 0.019$ & $2.820 \pm 0.019$ & $2.824 \pm 0.024$ \\
BHB   & $2.835 \pm 0.015$ & $2.900 \pm 0.020$ & $2.950 \pm 0.020$ & $2.980 \pm 0.020$ & $3.010 \pm 0.010$ & $3.024 \pm 0.025$ \\
DD2qG & $2.910 \pm 0.010$ & $3.000 \pm 0.020$ & $3.070 \pm 0.020$ & $3.110 \pm 0.020$ & $3.142 \pm 0.010$ & $3.152 \pm 0.021$ \\
\hline 
\hline
\end{tabular} 
\end{center} 
\end{table*}

\begin{figure*}[h!]
\centering
\includegraphics[width=0.49 \linewidth]{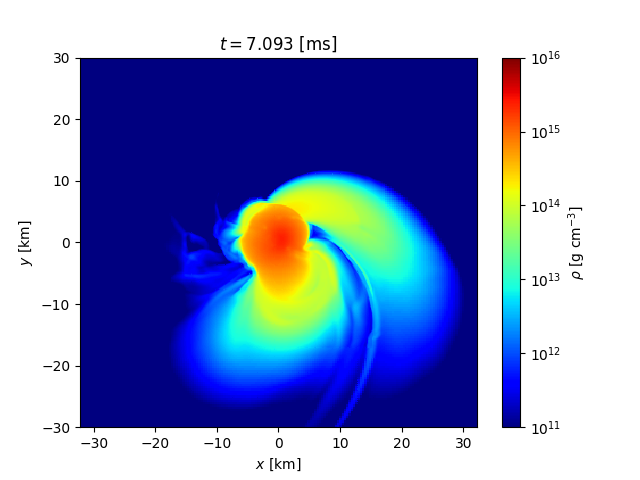}
\includegraphics[width=0.49 \linewidth]{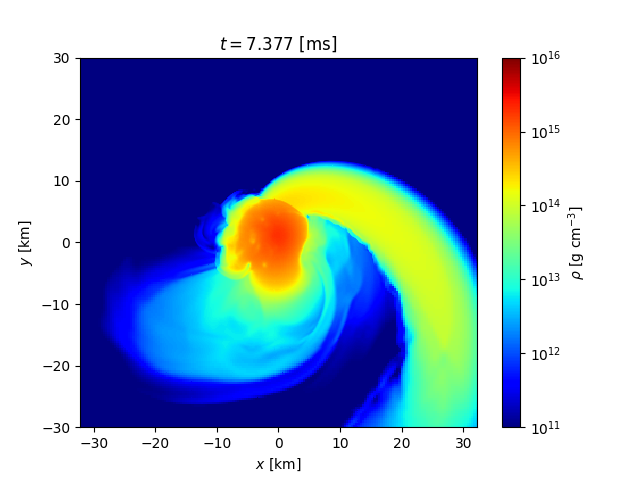}
\caption{Rest mass density in the orbital plane for two BNS merger simulations performed with the SFHo EOS, during the late inspiral phase and just prior to the prompt collapse of the remnant. The left panel refer to $q=0.85$, while the right one to $q=0.65$. In the former case, the tidal disruption of the secondary NS is marginal and the merger dynamics is driven by the core fusion, as testified by the ejection of shocked matter opposite to the tidal tail. In the latter case, the core of the secondary is significantly affected by tidal interaction and a large accretion stream flows onto the primary core.}
\label{fig: density snapshots different q's}
\end{figure*}

In \reffig{fig: density snapshots different q's}, we present the rest mass density on the orbital plane just after merger for a $M \gtrsim M_{\rm th}$ and for two SR simulations characterized by $q = 0.85$ (left panel) and $q = 0.65$ (right panel) to
highlight quantitative differences. Tidal interaction and deformation is present in
both cases, but with different features.
In the $q=0.85$ case, the two \ac{NS} cores are mostly preserved during the late 
inspiral and the subsequent remnant dynamics is governed by the core fusion process. 
This is visible in the left panel of \reffig{fig: density snapshots different q's} where, opposite to the tidal tail produced by the tidal stripping of secondary NS, a larger amount of shocked matter is expelled as consequence of the collision of the two cores. 
As $q$ decreases further below $\tilde{q}=0.725$, the secondary \ac{NS} is more significantly deformed during the last orbits and even its core gets affected. A significant accretion stream flows from the deformed core of the secondary toward the primary such that the debris of the secondary form an envelop that wraps the primary NS up \citep{Bauswein:2013yna,Dietrich:2015pxa,Dietrich:2016hky,Bernuzzi:2020txg,Bauswein:2020aag}.
This prominent change in the dynamics is visible in the right panel of of \reffig{fig: density snapshots different q's} where the tidal tail of the secondary is now much more relevant and the interaction between the two cores happens in form of an accretion stream rather than of the fusion of two comparable cores.

\subsection{Nuclear Incompressibility}

In \reftab{table: Kmax table}, we report some of the relevant properties of all the \acp{EOS} used to produce figure 4 of the main text. For each \ac{EOS}, in addition to the name, we report the mass and the radius of the heaviest \ac{NS}, $\Mmax,\Rmax$ respectively,  its central density, $\rhomaxTOV$, and the incompressibility at $\rhomaxTOV$, $\Kmax$. Overall, our sample consists of 66 cold, beta equilibrated \acp{EOS}. Among them, 12 include hyperons and 22 model a phase transition to quark matter.

\begin{table*}
\begin{center} 
\caption{Incompressibility at the maximum central density for each of the equations of state used in Figure 5 of this work. For each equation of state, in addition to the name, we report the properties of the heaviest neutron star, $(\Mmax,\Rmax)$, its central density, $\rhomaxTOV$, and the incompressibility at $\rhomaxTOV$, $\Kmax$.} 
\label{table: Kmax table} 
\begin{tabular}{ | c | c | c | c | c | c || c | c | c | c | c | c | } 
 \hline\hline 
   EOS    & $\Mmax$   & $\Rmax$      & $\rhomaxTOV$ & $\Kmax$ & Ref. 
 & EOS    & $\Mmax$   & $\Rmax$      & $\rhomaxTOV$ & $\Kmax$ & Ref. \\ 
            & $[\Msun]$ & $[{\rm km}]$ &   $[n_0]$  & [GeV]   &  &    
            & $[\Msun]$ & $[{\rm km}]$ & $[n_0]$    & [GeV]   &    \\
\hline 
 ALF2 & 1.99 & 11.31 & 6.05 & 6.0 & \cite{Alford:2004pf} 
& BA & 2.60 & 12.39 & 4.72 & 12.4 & \cite{Fattoyev:2017jql} \\
 BHB  & 2.10 & 11.59 & 5.94 & 8.2 & \cite{Banik:2014qja} 
& BL  & 2.10 & 10.50 & 7.18 & 17.2 & \cite{Bombaci:2018ksa,Logoteta:2020yxf} \\
 BSK20& 2.16 & 10.16 & 7.03 & 21.6 & \cite{Pearson:2011zz} 
& BSK21& 2.27 & 10.99 & 6.09 & 16.3 & \cite{Pearson:2012hz}  \\
 CMF  & 2.10 & 11.58 & 5.94 & 8.3 & \cite{Dexheimer:2008ax}  
& CMFy & 2.07 & 11.58 & 5.72 & 6.7 & \cite{Dexheimer:2015qha} \\
 CMFy1& 2.00 & 11.51 & 5.00 & 5.6 & \cite{Dexheimer:2015qha}  
& CMFy2& 2.07 & 11.42 & 6.13 & 8.6 & \cite{Dexheimer:2015qha} \\
 CMFy3& 2.07 & 11.47 & 6.02 & 8.4 & \cite{Dexheimer:2015qha} 
& DD2  & 2.42 & 11.90 & 5.26 & 12.8 & \cite{Typel:2009sy,Hempel:2011mk} \\ 
DD2F-RDF (1.1) & 2.14 & 10.21 & 7.17 & 16.5 & \cite{Bastian:2020unt} 
& DD2F-RDF (1.2) & 2.16 & 10.95 & 6.42 & 7.09 & \cite{Bastian:2020unt} \\
DD2F-RDF (1.3) & 2.04 & 10.37 & 7.28 & 10.74 & \cite{Bastian:2020unt} 
& DD2F-RDF (1.4) & 2.03 & 10.21 & 7.27 & 10.86 & \cite{Bastian:2020unt} \\
DD2F-RDF (1.5) & 2.04 & 10.34 & 7.27 & 10.63 & \cite{Bastian:2020unt} 
& DD2F-RDF (1.6) & 2.02 & 10.09 & 7.64 & 13.81 & \cite{Bastian:2020unt} \\
DD2F-RDF (1.7) & 2.13 & 10.81 & 6.67 & 8.37 & \cite{Bastian:2020unt} 
& DD2F-RDF (1.8) & 2.07 & 10.22 & 7.30 & 10.38 & \cite{Bastian:2020unt} \\
DD2F-RDF (1.9) & 2.17 & 10.78 & 6.44 & 6.69 & \cite{Bastian:2020unt} 
& DD2qG & 2.15 & 12.53 & 5.11 & 2.6 &  \cite{Kashyap:2021wzs} \\     
DDH  & 2.16 & 11.19 & 6.09 & 8.4 & \cite{Douchin:2001sv} 
& DDHy & 2.04 & 11.31 & 6.02 & 8.5 & \cite{Oertel:2014qza} \\ 
DD2hyp1& 2.00& 11.37 & 6.28 & 8.0 & \cite{Fortin:2017cvt} 
& DD2hyp2& 2.06& 11.65 & 5.90 & 7.9 & \cite{Fortin:2017cvt} \\ 
GM1  & 2.38 & 12.14 & 5.26 & 12.0 & \cite{Glendenning:1991es} 
& GM1y1& 2.29 & 11.90 & 5.26 & 10.1 & \cite{Oertel:2014qza} \\
GM1y2& 2.11 & 12.01 & 5.27 & 6.2 & \cite{Oertel:2014qza} 
&  H3 & 1.70 & 11.17 & 6.84 & 4.7 & \cite{Glendenning:1991es,Lackey:2005tk,Read:2009yp}  \\
H4   & 2.02 & 11.62 & 5.98 & 7.7 & \cite{Glendenning:1991es,Lackey:2005tk,Read:2009yp} 
&  HB   & 2.45 & 11.59 & 4.21 & 14.8 & \cite{Glendenning:1991es,Lackey:2005tk,Read:2009yp} \\
HOLO & 2.33 & 11.71 & 5.23 & 6.8 & \cite{Jokela:2020piw} 
&  IUFSU & 1.95 & 11.23 & 6.33 & 5.5 & \cite{Roca-Maza:2011alv} \\ 
LS220& 2.04 & 10.65 & 7.03 & 12.6 & \cite{Lattimer:1991nc} 
& MPA1& 2.47 & 11.28 & 5.49 & 23.6 & \cite{Muther:1987xaa} \\
 MS1  & 2.76 & 13.26 & 4.21 & 7.5 & \cite{Mueller:1996pm} 
& NL3 & 2.79 & 13.38 & 4.11 & 13.9 & \cite{Shen:2011kr} \\
OOS1 & 2.05 & 12.54 & 4.93 & 3.9 & \cite{Otto:2019zjy} 
&  OOS2 & 2.12 & 12.72 & 4.74 & 4.0 & \cite{Otto:2020hoz}  \\
OOS3 & 1.94 & 13.02 & 4.22 & 2.5 & \cite{Otto:2020hoz} 
&  QHC18 & 2.04 & 10.36 & 7.07 & 10.5 & \cite{Baym:2017whm} \\  
QHC19A&1.92 & 10.21 & 7.48 & 9.5 & \cite{Baym:2019iky} 
&  QHC19B&2.06 & 10.51 & 6.85 & 10.2 & \cite{Baym:2019iky} \\
QHC19C&2.18 & 10.73 & 6.43 & 11.0 & \cite{Baym:2019iky} 
&  QHC19D&2.27 & 10.83 & 6.21 & 11.9 & \cite{Baym:2019iky}  \\ 
SFHo & 2.05 & 10.32 & 7.18 & 13.7 & \cite{Steiner:2012rk} 
& SFHx & 2.13 & 10.74 & 6.59 & 10.3 & \cite{Steiner:2012rk} \\
 SKa  & 2.20 & 10.90 & 6.39 & 17.0 & \cite{Gulminelli:2015csa}  
& SKb  & 2.18 & 10.62 & 6.62 & 17.1 & \cite{Gulminelli:2015csa}  \\
 SkI6 & 2.19 & 10.76 & 6.47 & 16.0 & \cite{Gulminelli:2015csa} 
&  SLy4 & 2.05 & 10.01 & 6.47 & 13.6 & \cite{Chabanat:1997un} \\
 SRO(0)   & 2.20 & 10.74 & 6.51 & 15.8 & \cite{daSilvaSchneider:2017jpg,Kashyap:2021wzs} 
& SRO(2) & 2.16 & 11.06 & 6.40 & 9.1 & \cite{daSilvaSchneider:2017jpg,Kashyap:2021wzs} \\ 
 SRO(3) & 2.23 & 10.71 & 6.46 & 15.5 & \cite{daSilvaSchneider:2017jpg,Kashyap:2021wzs} 
& SRO(4) & 2.20 & 10.75 & 6.53 & 16.0 & \cite{daSilvaSchneider:2017jpg,Kashyap:2021wzs} \\ 
 SRO(5) & 2.19 & 10.65 & 6.64 & 18.9 &  \cite{daSilvaSchneider:2017jpg,Kashyap:2021wzs} 
& SRO(6) & 2.10 & 11.50 & 6.04 & 5.8 & \cite{daSilvaSchneider:2017jpg,Kashyap:2021wzs} \\ 
 SRO(7) & 2.22 & 10.72 & 6.53 & 16.3 & \cite{daSilvaSchneider:2017jpg,Kashyap:2021wzs} 
& SRO(8) & 2.18 & 10.73 & 6.67 & 16.8 & \cite{daSilvaSchneider:2017jpg,Kashyap:2021wzs} \\
  TM1  & 2.21 & 12.56 & 5.15 & 6.4 & \cite{Shen:1998gq,Hempel:2009mc} 
& TM1-2 & 2.25 & 12.23 & 5.28 & 6.8 & \cite{Logoteta.etal:2022} \\
 TM1-2q1 & 2.06 & 12.56 & 5.10 & 6.7 & \cite{Logoteta.etal:2022} 
& TM1-2q2 & 2.21 & 12.57 & 5.03 & 4.0 & \cite{Logoteta.etal:2022} \\ 
 TM1-2y & 1.98 & 12.27 & 5.42 & 4.7 & \cite{Logoteta.etal:2022} 
& TMA & 2.02 & 12.27 & 5.60 & 5.3 & \cite{Sugahara:1993wz} \\
\hline 
\hline 
\end{tabular} 
\end{center} 
\end{table*}

\subsection{Alternative analysis and comparison with exisiting fits}

\begin{figure*}
\centering
\includegraphics[width=0.49 \linewidth]{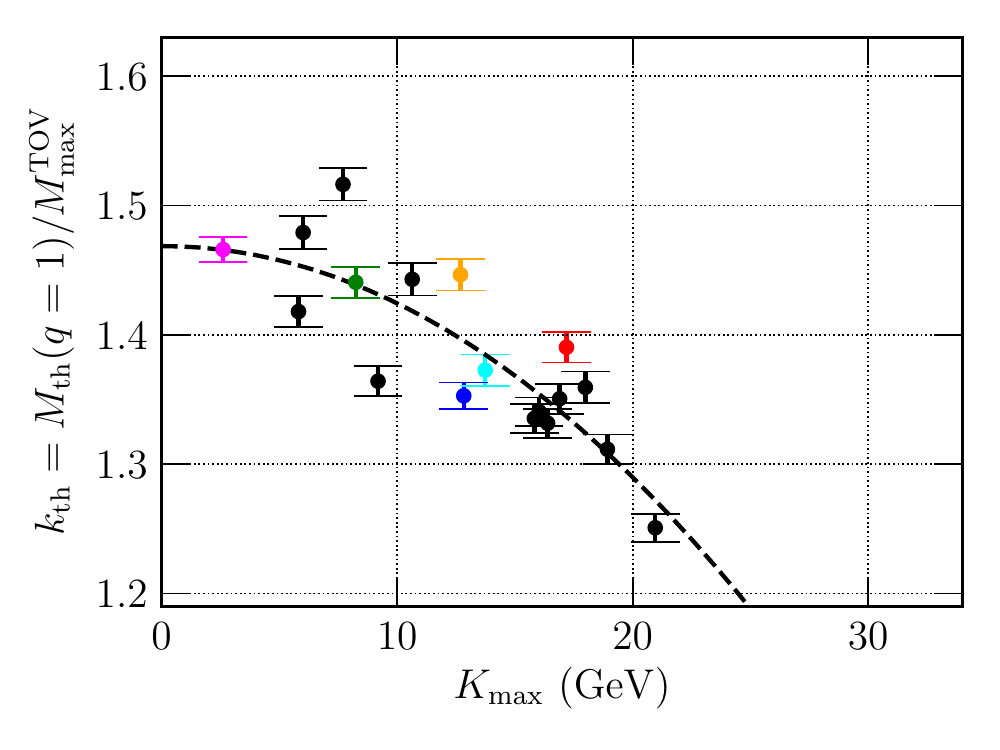}
\includegraphics[width=0.49 \linewidth]{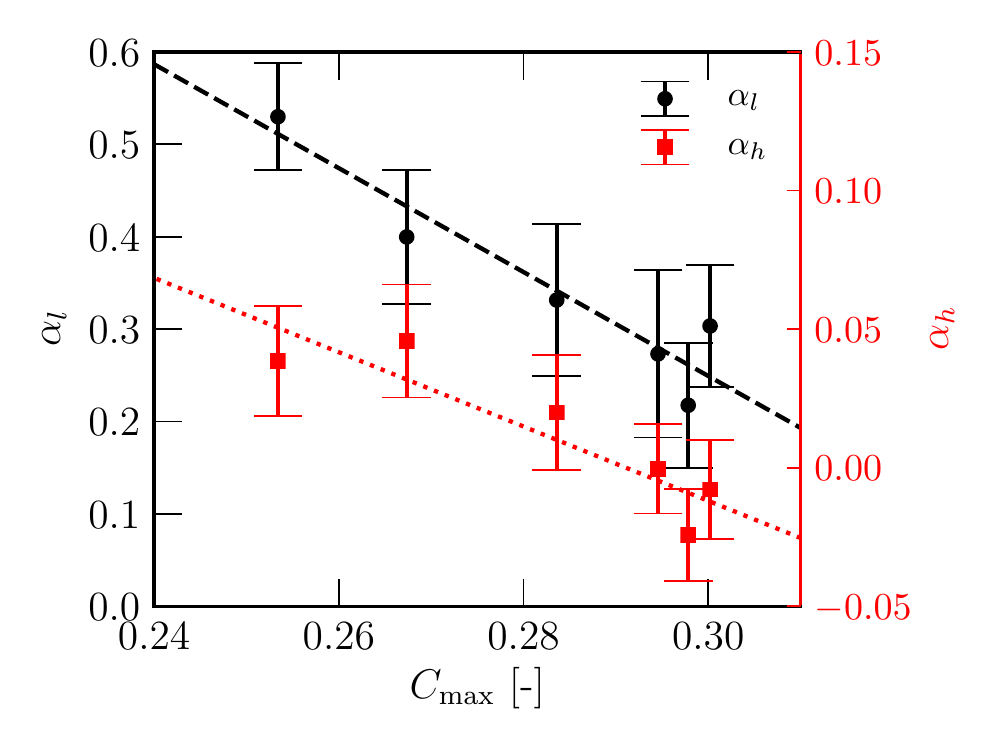}
\caption{Left: $k_{\rm th} \equiv \mthpc(q=1)/\Mmax$ as a function of $\Kmax$ from the $q=1$ results reported in \cite{Kashyap:2021wzs}, for the \acp{EOS} for which $\Mmax > 1.97 \Msun$ 
The dashed line correspond to the power-law fit described in the text. The colored points corresponds to the six \acp{EOS} used in this work. Right: Slope parameters, $\alpha_{h,l}$, presented in the main text (see equations 2 and 3), as a function of the maximum compactness, $\Cmax$. The dashed and dotted lines represent a linear fit with respect to $\Cmax$.}
\label{fig: alternative fits}
\end{figure*}

\begin{figure*}
\centering
\includegraphics[width=0.55 \linewidth]{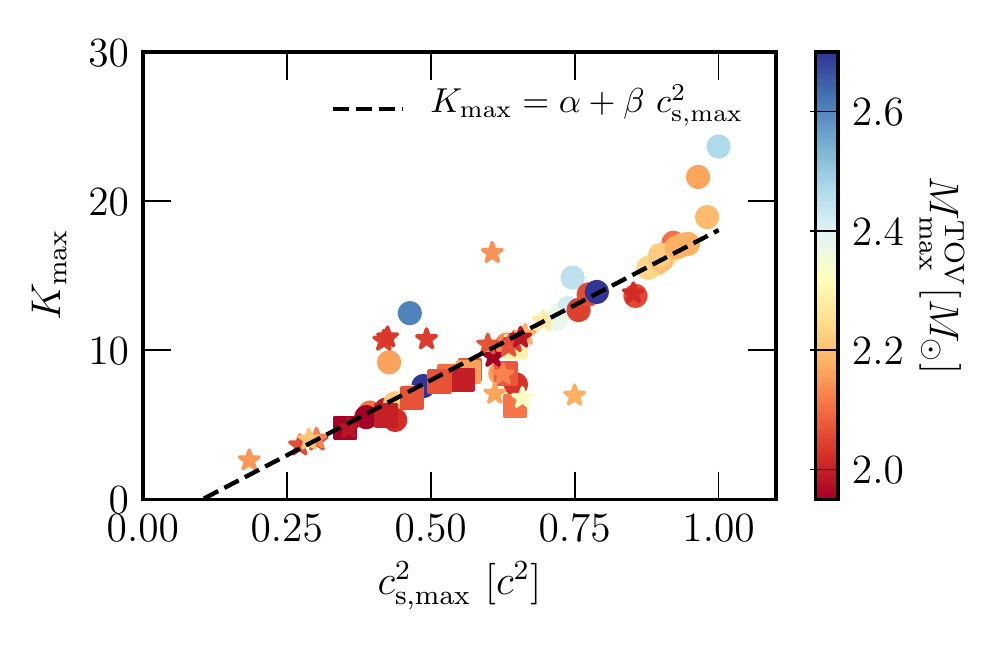}
\includegraphics[width=0.44 \linewidth]{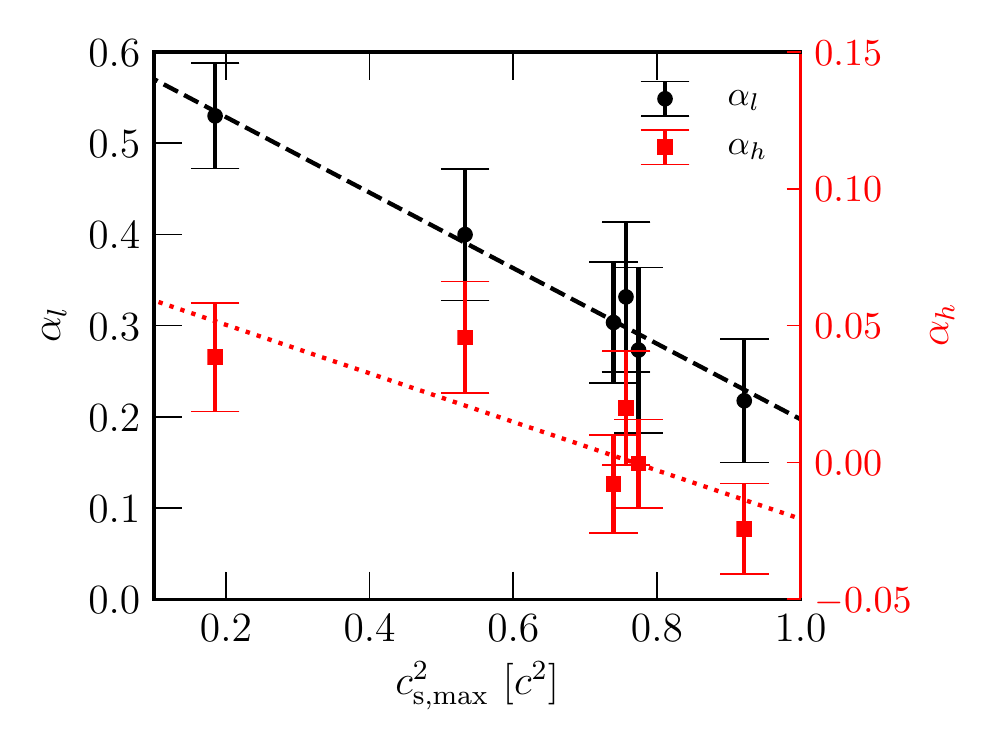}
\caption{Left: $\Kmax$ as a function of $c_{\rm s,max}^2$ for the \acp{EOS} presented in \reftab{table: Kmax table}. The dashed line correspond to a linear fit. 
Right: Slope parameters, $\alpha_{h,l}$, presented in the main text (see equations 2 and 3), as a function of the square of the speed of sound at $\rhomaxTOV$, $c_{\rm s,max}^2$. The dashed and dotted lines represent a linear fit with respect to $c_{\rm s,max}^2$.}
\label{fig: alternative fits cs}
\end{figure*}

As highlighted in the main text, the relation between $\Cmax$ and 
$\Kmax$ implied by figure 4 suggests the possibility of repeating the \ac{PC} analysis in terms of a single variable, either $\Cmax$ or $\Kmax$.
We first consider the $q=1$ data presented in Ref.~\citep{Kashyap:2021wzs} for which
the linear fit $\mthpc(q=1)/\Mmax = a \Cmax + b$ was proposed 
\citep[see e.g.][]{Bauswein:2013jpa}. By inserting the ansatz 
$\Kmax = \alpha \left(\Cmax \right)^{1/\delta}$, we obtain 
$\mthpc(q=1)/\Mmax = a' \left( \Kmax/ 1{\rm GeV}\right)^{d'} + b'$. 
Standard least square methods yield to 
$a'=\left( -4.3 \pm 3.5 \right) \times 10^{-4} $, $b'= 1.469 \pm 0.009 $ and 
$d'=2.01 \pm 0.26 $. The points and the result of the fit are reported in the left panel of \reffig{fig: alternative fits}. However, the fit in $\Cmax$ should be 
preferred since it provides a tighter correlation and it uses one fitting 
parameter less.

Additionally, we repeat our $q \neq 1$ analysis in terms of $\Cmax$. In the right panel of \reffig{fig: alternative fits}, we present fits for the $\alpha_{h,l}$
coefficients introduced in equation 2 of the main text in terms of $\Cmax$. 
The results are:
\begin{eqnarray}
    \alpha_l &=& - (5.62 \pm 0.92) \Cmax + (1.94 \pm 0.26)  \, , \nonumber \\
    \alpha_h &=& - (1.34 \pm 0.32) \Cmax + (0.390 \pm 0.091)  \, .
    \label{eq: fit alternative coefficients}
\end{eqnarray}
The quality of the fits (estimated for example by the adjusted coefficient of determination, $R^2_{\rm adj}$) are comparable, but overall slightly worse than the ones obtained in the main text for $\Kmax$: 
$R^2_{\rm adj}$ moves from 0.70 and 0.99 for the fits in $\Kmax$ to
0.76 and 0.88 for the fits in $\Cmax$, for the high- and low-$q$ fits, respectively.

Finally, given the analogy between the definition of $K$ (equation 1 of the main text) and the square of the speed of sound of \ac{NS} matter, $c_{\rm s}^2 = \left. \partial P/\partial \epsilon \right|_{T=0,\delta_{\rm eq}}$, where $P$ and $\epsilon$ are the pressure and the density of internal energy, and $\delta_{\rm eq}$ is the symmetry parameter for \ac{NS} matter in neutrinoless weak equilibrium, we investigate also the relation between $\Kmax$ and $c_{\rm s,max}^2$, defined as the square of the speed of sound at $\rhomaxTOV$. In the left panel of \reffig{fig: alternative fits cs} we present $\Kmax$ as a function of $c_{\rm s,max}^2$ for all the \acp{EOS} in \reftab{table: Kmax table}. With the exception of a few outliers, a linear relation between $\Kmax$ and $c_{\rm s,max}^2$ holds. 
We perform a linear interpolation, $\Kmax = \alpha + \beta~\left( c_{\rm s,max}/c\right)^2$, finding
$\alpha = \left(-2.09 \pm 0.86 \right){\rm GeV} $ and $\beta= \left( 20.1 \pm 1.3 \right){\rm GeV} $, with $R_{\rm adj}^2 = 0.78$.
For $c_{\rm s,max}^2$ approaching unity, a possible divergence or an upper limit for $\Kmax$ are suggested. However, no firm conclusions can be drawn from the paucity of models and from the intrinsic limitation of non-relativistic nucleonic \acp{EOS} in modeling such a regime.
Motivated by this result, we repeat our $q \neq 1$ analysis also in terms of $c_{\rm s,max}^2$. In the right panel of \reffig{fig: alternative fits cs}, we present fits for the $\alpha_{h,l}$ coefficients introduced in equation 2 of the main text as a function of $c_{\rm s,max}^2$. 
The results are:
\begin{eqnarray}
    \alpha_l &=& - (0.414 \pm 0.027) c_{\rm s,max}^2 + (0.611 \pm 0.018)  \, , \nonumber \\
    \alpha_h &=& - (0.089 \pm 0.030) c_{\rm s,max}^2 + (0.068 \pm 0.022)  \, .
    \label{eq: fit alternative coefficients cs2}
\end{eqnarray}
Also in this case, the quality of the fits are comparable, but overall slightly worse, than the ones obtained in the main text for $\Kmax$: 
$R^2_{\rm adj}$ is 0.60 and 0.95 for the high- and low-$q$ fits, respectively.

\begin{figure*}[h!]
\centering
\includegraphics[width=\linewidth]{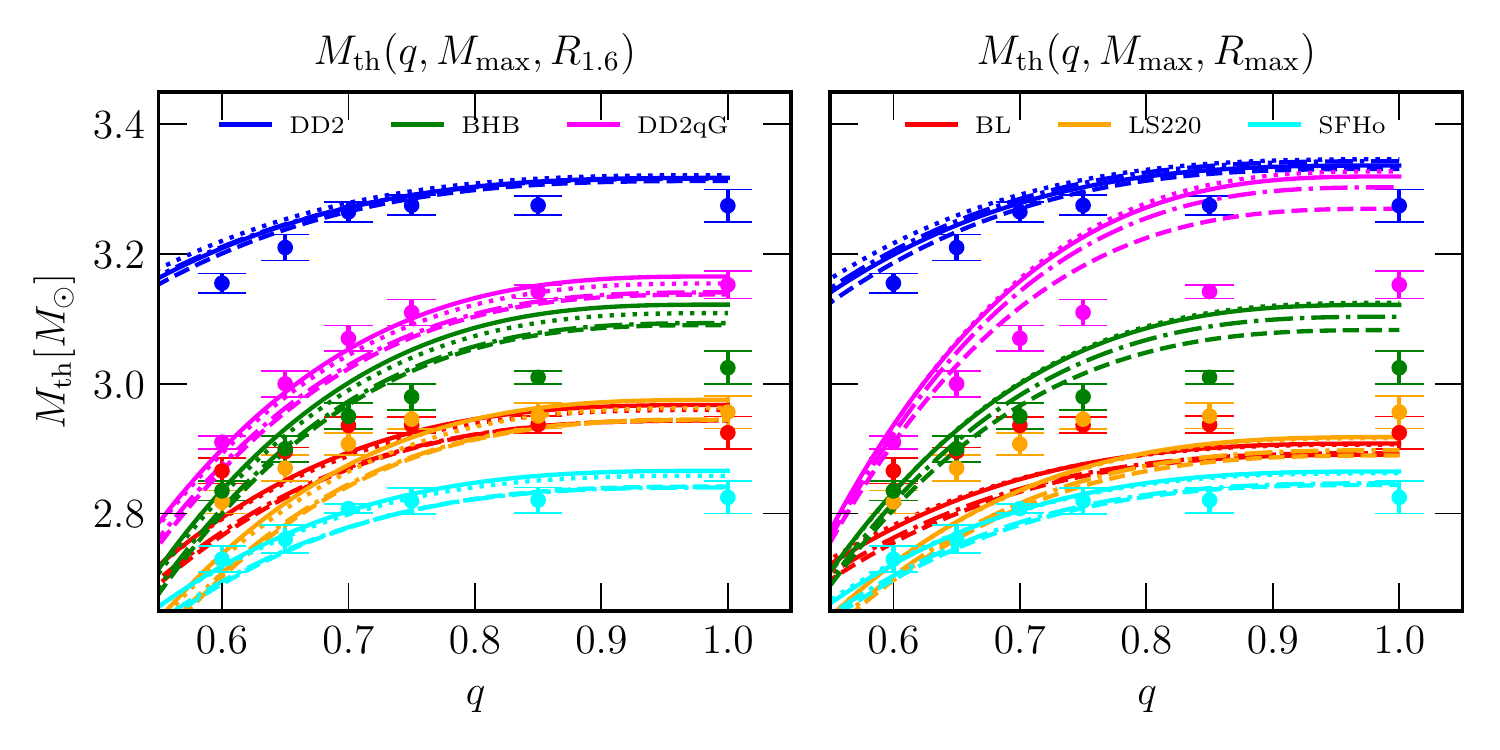}
\caption{Comparison between our results (dots) and the $q$-dependent fit results presented in \cite{Bauswein:2020xlt}, equation 10 and table VI. Different colors refer to the six \ac{EOS} used in our work, while different lines correspond to different \ac{EOS} samples employed in Table IV of \cite{Bauswein:2020xlt}. In particular, we include the baseline "b" sample (solid), the "b+h" sample (dashed), the "b+e" sample (dotted) and the "b+h+e" sample (dash-dotted).}
\label{fig: comparison with Bauswein fits}
\end{figure*}

The study of the behavior of $\mthpc(q)$ has been the subject of several recent works, that provided useful fits in different variables and according to different ansatz.
Tootle {\it et al.} \cite{Tootle:2021umi} and Kolsch {\it et al.} \cite{Kolsch:2021lub} used a definition of \ac{PC} that is not equivalent to ours. Despite providing similar results, the average and the maximum variations between the two methods reported by Ref.~\cite{Tootle:2021umi} are larger than our numerical errors and comparable to the overall variation we are interested to model. Moreover, Ref.~\cite{Tootle:2021umi} studied the possible existence
of a quasi-universal relation, while here we focus on a \ac{EOS} dependent relation. Thus, we abstain from a detailed comparison with the results reported in these two works.
On the other hand, our definition of \ac{PC} is equivalent to the one used in \citep{Bauswein:2020xlt} and thus we can compare our results with the polynomial fits discussed in \citep{Bauswein:2020xlt} and presented in their equation 10. 
In \reffig{fig: comparison with Bauswein fits}, we compare our results
with all the sets of fits ("b", "b+h", "b+e", "b+h+e") reported in their
table VI for $\mthpc = \mthpc(q,\Mmax,R_{1.6})$ and $\mthpc = \mthpc(q,\Mmax,\Rmax)$.
We note that the fit in $(q,\Mmax,R_{1.6})$ predicts results that are,
on average, closer to ours than the fit in $\mthpc(q,\Mmax,\Rmax)$, despite the fact that the latter shares the same set of independent variables. However, for most of the \acp{EOS}, differences larger than our numerical uncertainties and than the fit residuals reported in \cite{Bauswein:2020xlt} can be observed between the fits and our results in at least a portion of the relevant $q$ interval. In particular, both the fits are not able 
to reproduce the slight increase in $\mthpc(q)$ for $q \lesssim 1$ for the BLh EOS, as also observed by Kolsch {\it et al.} for other \acp{EOS} \cite{Kolsch:2021lub}.

\end{document}